\documentclass[onecolumn,10pt]{IEEEtran}
\IEEEoverridecommandlockouts
\ifCLASSINFOpdf
\usepackage{graphicx}
\usepackage{epstopdf}
\usepackage{float}
\else
\fi
\usepackage{amsfonts}
\usepackage{amssymb}
\usepackage{amsmath}
\usepackage{accents}
\usepackage{multirow}
\usepackage{array}
\usepackage{pgfplots}
\usepackage{booktabs}
\usepackage{multirow}
\usepackage{epstopdf,epsfig}

\definecolor{mycolor1}{rgb}{0.60000,0.20000,0.00000}%
\definecolor{mycolor2}{rgb}{0.00000,0.20000,0.60000}%
\definecolor{mycolor3}{rgb}{0.00000,0.40000,0.80000}%
\definecolor{mycolor4}{rgb}{0.00000,0.60000,1.00000}%
\definecolor{mycolor5}{rgb}{1.00000,0.00000,0.20000}%
\definecolor{mycolor6}{rgb}{1.00000,0.20000,0.40000}%

\begin{document}

\title{Study of Rate-Splitting Techniques with Block Diagonalization for Multiuser MIMO Systems}

\author{Andre R. Flores $^1$ and
and Rodrigo C. de Lamare $^1$,$^2$ \\
$^1$ ~Centre for Telecommunications Studies, Pontifical Catholic
University of Rio de Janeiro, Brazil \\

$^2$ ~Department of Electronic Engineering, University of York,
United Kingdom \\
Emails: andre.flores@cetuc.puc-rio.br,
delamare@cetuc.puc-rio.br
\thanks{This work is partly funded by the CNPq and FAPERJ Brazilian agencies.}\vspace{-0.75em}}

\maketitle

\begin{abstract}
In this work, we investigate Block Diagonalization (BD) techniques for multiuser multiple-antenna systems using rate-splitting (RS) multiple access. In RS multiple access the messages of the users are split into a common part and a private part in order to mitigate multiuser interference. We present the system model for a RS multiple access system operating in a broadcast channel scenario where the receivers are equipped with multiple antennas. We also develop linear precoders based on BD for the RS multiple access systems along with combining techniques, such as the min-max criterion and the maximum ratio combining criterion, to enhance the common rate. Closed-form expressions to describe the sum rate performance of the proposed scheme are also derived. The performance of the system is evaluated via simulations considering imperfect channel state information at the transmitter. The results show that the proposed schemes outperform conventional linear precoding methods.

\end{abstract}
\vspace{0.1cm}
\begin{IEEEkeywords}
Multiple-antenna systems, ergodic sum-rate, rate-splitting, block diagonalization .
\end{IEEEkeywords}
\IEEEpeerreviewmaketitle

\section{Introduction}
Multiple-Input Multiple-Output (MIMO) technology enhances the
information and error rates performance of wireless communications
systems by exploiting multipath propagation through multiple
transmit and receive antennas \cite{Telatar1999}. Modern wireless
communications systems deal with multiple users distributed
geographically, making multiuser MIMO (MU-MIMO) the focus of many
research works over the last decade. Among the key problems of
MU-MIMO are  the multi-user interference (MUI) and acquisition of
channel state information at the transmitter (CSIT), which can
decrease dramatically the overall system performance
\cite{Lu2014,mmimo_delamare}. In order to deal with MUI, many
transmit processing techniques have been proposed in the literature
\cite{Joham2005,Spencer2004,ZuLamare2012,mmimo_mismatch}. Most of
these techniques rely on the quality of CSIT. Nevertheless,
obtaining highly accurate CSIT in practice is still an open problem
\cite{JoudehClerckx2016}.

Rate-splitting (RS) multiple access schemes, proposed in \cite{Han1981}, have been
adopted in the last years as a promising approach to enhance the
sum-rate performance of MIMO systems working under imperfect CSIT.
Basically, RS splits the data before transmission
into a common stream and private streams. The common stream should
be decoded by all users. In contrast, the private streams are decoded
only by its corresponding user. The main advantage of RS schemes is
that they can adjust the content and the power of the common message
in order to partially decode interference and partially treat interference
as noise. RS has the ability to control how much interference should
be decoded through the common message and how much should be treated
as noise \cite{Maoinpress}.

In the literature, RS has been used in conjunction with several linear
precoding techniques \cite{JoudehClerckx2016,Hao2015}, working under
perfect and imperfect CSIT assumption. RS with non-linear precoders
has been studied in \cite{Flores2018}. Other scenarios of interest,
such as massive MIMO and MISO networks with RS have been considered
in \cite{Dai2016} and \cite{Hao2017}, respectively. RS has also been
studied for robust transmission under bounded CSIT errors
\cite{Joudeh2016a}. However, most of the works on RS consider only
multiple-input single-output (MISO) scenarios and zero-forcing and minimum mean-squared error (MMSE) channel inversion-type precoders. MIMO scenarios have been studied in \cite{HaoB.Clerckx2017} from a DoF perspective. In \cite{Kolawole2018} a MIMO RS architecture has been proposed for millimetre waves using a ZF precoding. However, the design of linear precoders for RS schemes has not considered Block Diagonalization (BD) type linear precoders, which have the potential to significantly enhance the sum rate performance of ZF and MMSE linear precoders.

In this paper, we generalize linearly precoded RS multiple access schemes
to MU-MIMO systems where the users are equipped with multiple
antennas. We present BD linear precoder for RS multiple access schemes in MU-MIMO
systems. Furthermore, we propose techniques based on the min-max and the maximum ratio combining criteria to enhance the common rate
at the receivers with multiple antennas. The performance of
the proposed schemes is evaluated using the sum rate figure of merit in a
Broadcast Channel (BC) under imperfect CSIT assumption.

The rest of this paper is organized as follows. Section II presents
the mathematical model of the system and briefly reviews linear
precoding techniques. Section III describes the proposed combining
strategies to compute the common rate. Section
IV present the analysis of the sum rate performance. Section V shows the simulation results.
Finally, Section VI draws the conclusions of this work.

Matrices are denoted by boldface uppercase letters, whereas boldface
lower case letters denote column vectors. Standard letters represent
scalars. The supperscripts $\left(\cdot\right)^{\text{T}}$ and
$\left(\cdot\right)^{H}$ are the transpose and Hermitian operators
respectively. The cardinality operator is given by
$\text{card}\left(\cdot\right)$ and
$\text{diag}\left(\mathbf{a}\right)$ represents a diagonal matrix
with the entries of vector $\mathbf{a}$ in the main diagonal. The trace of a matrix is denoted by $\text{tr}\left(\cdot\right)$.
$\mathbb{E}\left[\cdot\right]$ stands for the expectation operator,
$\lVert\cdot \rVert$ for the Euclidean norm
and  $\odot$ for the Hadamard product.

\section{System Model and Linear Precoding}

We consider a MIMO BC with $K$ users, where user $k$ is equipped
with $N_k$ antennas. The total number of receive antennas is then
given by $N_r=\sum_{k=1}^{K}N_k$. The number of antennas at the
transmitter is denoted by $N_t$ and remains in the range $ N_t \geq
K \geq 2$. The
group of data streams intended for user $k$ form a set denoted by
$\mathcal{B}_k$. Then, the $i$th stream sent to the $j$th user is
expressed by $B_{ij}$. The total number
of transmitted data streams is $B=\sum_{k=1}^{K} M_k$ with
$M_k=\text{card}\left(\mathcal{B}_k\right)$. The model satisfies the
transmit power constraint
$\mathbb{E}\left[\lVert\mathbf{x}\rVert^2\right]\leq E_{tr}$, where the vector
$\mathbf{x}$ represents the transmitted signal and
$E_{tr}$ denotes the total transmit power.

The system employs the RS scheme, which splits the messages into a
common part and a private part
\cite{Clerckx2016,Han1981}. Since we focus on the sum rate analysis, it suffices to  consider that only one stream is split. 
The common part is encoded into one 
common stream and the private parts into $B$ private streams. The receivers
share a codebook since the common message has to be decoded by all
the users with zero error probability. In contrast, each private
stream is decoded only by its corresponding user. This means that
each receiver must decode $M_k+1$ 
data streams, namely the common stream (decoded by all but intended to only one user) and 
a set of private streams (decoded by its respective user). This is
possible if we apply successive interference cancellation (SIC)
techniques
\cite{vblast,isic1999,itic,spadf,mbsic,mfsic,mbdf,did,bf_idd,idd1bit}.
The common stream is first decoded using SIC and all private
messages are considered as interference and treated as noise. At the
end, the message sent via the private streams is decoded. The
strength of RS is its ability to adjust the content and the power of
the common message to control how much interference should be
decoded by all users (through the common message) and how much
interference is treated as noise.

The information sequences in the data streams are modulated and then the splitting process is performed over the message, resulting in a vector of data symbols $\mathbf{s}^{\left(\text{RS}\right)} \in \mathbb{C}^{B+1}$. Specifically, the vector of the transmitted symbols is given by
$\mathbf{s}^{\left(\text{RS}\right)}=\left[s_c,\mathbf{s}_1^{\text{T}},\mathbf{s}_2^{\text{T}},\dots,\mathbf{s}_K^{\text{T}}\right]^{\text{T}}$,
where $s_c$ is used to designate the symbol of the common stream and $\mathbf{s}_k$ contains the $M_k$ private streams of the $k$th user. We assume uncorrelated symbols with zero mean and covariance
matrix equal to $\mathbf{R_{ss}}=\mathbf{I}$. The transmitter
processes the symbols using a linear precoder, which maps the
symbols to the transmit antennas. A common precoder, $\mathbf{p}_c \in \mathbb{C}^{N_t}$ is introduced in order to map the common symbols to the transmit antennas. Then, the precoder is given by $\mathbf{P}^{\left(\text{RS}\right)}=\left[\mathbf{p}_c,\mathbf{P}_1,\mathbf{P}_2,\dots,\mathbf{P}_K\right]$
, where $\mathbf{P}_k\in \mathbb{C}^{N_t\times M_k}$ is the precoder of the $k$th user.  Moreover, the vector $\mathbf{p}_k$ denotes the $k$th column of matrix $\mathbf{P}^{\left(\text{RS}\right)}$.
 The transmitted signal is expressed by
\begin{align}
\mathbf{x}=&\mathbf{P}^{\left(\text{RS}\right)}\mathbf{A}^{\left(\text{RS}\right)}\mathbf{s}^{\left(\text{RS}\right)}\nonumber\\
=&a_c s_c\mathbf{p}_c +\sum_{i=1}^K\mathbf{P}_i\text{diag}\left(\mathbf{a}_i\right)\mathbf{s}_i,\label{transmit signal}
\end{align}
where $\mathbf{A}^{\left(\text{RS}\right)} \in \mathbb{R}^{\left(B+1\right)\times \left(B+1\right)}$ is a general diagonal power loading matrix and the vector $\mathbf{a}^{\left(\text{RS}\right)}=\left[a_c,\mathbf{a}_1^{\text{T}},\mathbf{a}_2^{\text{T}},\dots,\mathbf{a}_K^{\text{T}}\right]^{\text{T}}\in \mathbb{R}^{\left(B+1\right)}$ consists of the coefficients in the main diagonal of the matrix $\mathbf{A}^{\left(\text{RS}\right)}$. The vector $\mathbf{a}_k \in \mathbb{R}^{M_k}$ contains the power assigned to the $M_k$ symbols intended for the $k$th user. The coefficient $a_c$ denotes the power distributed to the common message. In other words, the total transmit power is allocated partially to the common and private streams. The transmit power constraint is expressed by
$\text{tr}\left(\mathbf{P}^{\left(\text{RS}\right)}\text{diag}\left(\mathbf{a}^{\left(\text{RS}\right)}\odot\mathbf{a}^{\left(\text{RS}\right)}\right)\mathbf{P}^{\left(\text{RS}\right)H}\right)\leq
E_{tr}$.
Assuming normalized precoders, the transmit power constraint is reduced to $\lvert a_c \rvert^2 + \sum_{k=1}^B{\lvert a_k\rvert^2}\leq E_{tr}.$

After the precoding operation, the data are sent to the receiver over
the channel $\mathbf{H}=\mathbf{\hat{H}}+\mathbf{\tilde{H}} \in \mathbb{C}^{N_r\times N_t}$, where each
coefficient $h_{ij}$ in the channel matrix $\mathbf{H}$ represents
the link between the $j$th transmit antenna and the $i$th receive
antenna. The matrix $\mathbf{\hat{H}}$ represents the estimate of the channel and the matrix $\mathbf{\tilde{H}}$ takes into account the quality of the channel estimate by modelling the error produced by the estimation procedure. The channel of the $k$th user is given by $\mathbf{H}_k\in\mathbb{C}^{N_k\times N_t}$. It follows that $\mathbf{H}= [{\mathbf H}_1^{H} \ldots {\mathbf H}_k^{H}
\ldots {\mathbf H}_K^{H}]^H$. The vector $\mathbf{h}_j^{\text{T}} \in \mathbb{C}^{N_t}$ represents the $j$th row of matrix $\mathbf{H}$ . For simplicity, we consider a flat fading
channel which remains fixed during a transmission block.

The received signal obtained following the model in \eqref{transmit signal} is
given by
\begin{equation}
\mathbf{y}=\mathbf{G}\left(\mathbf{H}\mathbf{P}^{\left(\text{RS}\right)}\text{diag}\left(\mathbf{a^{\left(\text{RS}\right)}}\right)\mathbf{s^{\left(\text{RS}\right)}}+\mathbf{n}\right),\label{General Receive vector}
\end{equation}
where $\mathbf{G} \in \mathbb{C}^{B\times N_r}$ is a block diagonal receive filter since there is no cooperation between users. Moreover, the matrix $\mathbf{G}_k \in \mathbb{C}^{M_k\times N_k}$ denotes the receive filter of the $k$th user.
The vector $\mathbf{n}\in\mathbb{C}^{N_r\times 1}$ is the additive noise
modelled as a circularly symmetric complex Gaussian random vector,
i.e., $\mathbf{n}\sim
\mathcal{CN}\left(\mathbf{0},\mathbf{R_{nn}}\right)$. Without loss of
generality, we will consider that the noise is uncorrelated and has
the same statistical properties at each antenna, i.e.,
$\sigma_{n,i}^2=\sigma_{n,j}^2=\sigma_n^2$, $\forall i,j$, reducing
the covariance matrix to $\mathbf{R_{nn}}=\sigma_n^2\mathbf{I}$. The SNR is defined as $\text{SNR}\triangleq E_{tr}/ \sigma_n^2$.

Given a channel state and considering an imperfect CSIT scenario, the received signal at the $k$th terminal can
be written as
\begin{align}
\mathbf{y}_k=&\overbrace{a_c s_c \mathbf{G}_k\mathbf{H}_k\mathbf{p}_c}^{\text{Common Message}} +\overbrace{\sum_{i\in\mathcal{B}_k}a_i s_i\mathbf{G}_k\mathbf{H}_k\mathbf{p}_i}^{\text{User Data}}\nonumber\\
&+\overbrace{\sum_{\substack{j=1\\j\notin\mathcal{B}_k}}^{B} a_js_j\mathbf{G}_k\mathbf{H}_k\mathbf{p}_j}^{\text{Multi-User Interference}}+\mathbf{G}_k\mathbf{n}_k.\label{Received signal RS user k}
\end{align}

Let us define the matrix $\mathbf{F}^H=\mathbf{G}\mathbf{H}$ and the matrix $\mathbf{\hat{F}}^H=\mathbf{G}\mathbf{\hat{H}}$. The mean power of the $l$th received stream at user $k$ can be expressed as follows:
\begin{align}
\mathbb{E}\left[\lvert y_{lk}\rvert^2\right]=&a_c^2  \lvert\mathbf{f}_l^H\mathbf{p}_c\rvert^2 +\sum_{i\in\mathcal{B}_k}a_i^2 \lvert\mathbf{f}_l^H\mathbf{p}_i\rvert^2\nonumber\\
&+\sum_{\substack{j=1\\j\notin\mathcal{B}_k}}^B a_j^2\lvert\mathbf{f}_l^H\mathbf{p}_j\rvert^2+\lVert\mathbf{g}_{l}\rVert^2\sigma_n^2.\label{Received signal RS antenna k}
\end{align}

When working under a perfect CSIT scenario, the term $\mathbf{\tilde{H}}$ is reduced to zero and equations \eqref{Received signal RS user k} and \eqref{Received signal RS antenna k} remain the same with $\mathbf{H}=\mathbf{\hat{H}}$.

Suppose that we allocate no power to the common stream, i.e., we set $a_c=0$. In such cases the model represents a conventional MU-MIMO system where no RS is performed. It turns out that the model established is a general framework and conventional MU-MIMO can be seen as a particular case where no power is assigned to the common stream.

In what follows we review the main concepts about the ZF  and BD precoding techniques. We consider that the precoder of the private stream is defined as $\mathbf{P}=\left[\mathbf{P}_1,\mathbf{P}_2,\dots,\mathbf{P}_K\right]$.
\subsection{Linear ZF Precoding}

The ZF precoder \cite{Joham2005} defined by
\begin{equation}
\mathbf{P}^{\left(\text{ZF}\right)}=\beta^{\left(\text{ZF}\right)}\mathbf{H}^H\left(\mathbf{H}\mathbf{H}^H\right)^{-1},\label{ZFprecoder}
\end{equation}
where $\beta^{\left(\text{ZF}\right)}$ is a scaling factor introduced to satisfy the transmit power constraint that is defined as
\begin{equation}
\beta^{\left(\text{ZF}\right)}=\sqrt{\frac{E_{tr}}{\text{tr}\left(\left(\mathbf{H}\mathbf{H}^H\right)^{-1}\mathbf{R}_{ss}\right)}}
\end{equation}

\subsection{Linear BD Precoding}

BD precoding has been proposed in \cite{Spencer2004} and
\cite{Choi2004} for wireless communications systems, and further
studied in \cite{SungLeeLee2009,Keke2013,Zhangothers2017} due its
potential to increase the sum rate performance. This technique is
based on Singular Value Decomposition (SVD). The precoding matrix
for the $k$th user can be written in two parts as follows:
\begin{equation}
\mathbf{P}_{k}^{\left(\text{BD}\right)}=\mathbf{P}^a_k\mathbf{P}^b_k.\label{BDprecoder}
\end{equation}
The filter $\mathbf{P}^a_k$ is used to completely eliminate the MUI, whereas the filter $\mathbf{P}^b_k$ allows parallel symbol detection.
Let us form the matrix $\mathbf{\bar{H}}_k$ by excluding the channel matrix of the $k$th user, i.e. $\mathbf{\bar{H}}_k=\left[\mathbf{H}_1^H \ldots \mathbf{H}_{k-1}^H ~\mathbf{H}_{k+1}^H \ldots \mathbf{H}_K^H\right]$. By using SVD we get $\mathbf{\bar{H}}_k=\mathbf{\bar{U}}_k\mathbf{\bar{\Psi}}_k\left[\mathbf{\bar{V}}_k^{\left(1\right)} \mathbf{\bar{V}}_k^{\left(0\right)}\right]^H$. The matrix $\mathbf{\bar{V}}_k=\left[\mathbf{\bar{V}}_k^{\left(1\right)} \mathbf{\bar{V}}_k^{\left(0\right)}\right]^H$ is a unitary matrix with dimensions $N_t\times N_t$. Let us suppose that the rank of $\mathbf{\bar{H}}_k$ is given by $\bar{L}_k$. The vector $\mathbf{\bar{V}}^{\left(0\right)}_k$ contains the last $N_t-\bar{L}_k$ singular vectors and forms an orthogonal basis for the null space of $\mathbf{\bar{H}}_k$. Therefore, we can set the first part of the precoder to
\begin{equation}
\mathbf{P}^a_k=\mathbf{\bar{V}}_k^{\left(0\right)}.
\end{equation}
The first precoder separates the MU-MIMO channel into $K$ parallel
independent channels. Consider the effective channel matrix defined
as $\underaccent{\ddot}{\mathbf{H}}_{k}=\mathbf{H}_k\mathbf{P}^a_k$.
Performing a second SVD over  the effective channel
$\underaccent{\ddot}{\mathbf{H}}_{k}$, i.e.
$\underaccent{\ddot}{\mathbf{H}}_{k}=\underaccent{\ddot}{\mathbf{U}}_k
\underaccent{\ddot}{\mathbf{\Psi}}_k\left[\underaccent{\ddot}{\mathbf{V}}_k^{\left(1\right)}
\underaccent{\ddot}{\mathbf{V}}_k^{\left(0\right)}\right]^H$ we
obtain the second precoder and the receive filter as given by
\begin{align}
\mathbf{P}^b_k=&\underaccent{\ddot}{\mathbf{V}}_k^{\left(1\right)}, & \mathbf{G}^{\left(\text{BD}\right)}_k=&\underaccent{\ddot}{\mathbf{U}}_k^H.
\end{align}
The matrices $\mathbf{P}_k^{b}$ and
$\mathbf{G}^{\left(\text{BD}\right)}_k$ allow us to perform
symbol-by-symbol detection.

\section{RS Common and Private Rates}

The sum rate performance of a system employing an RS architecture
and assuming Gaussian signalling consists of a common rate $R_c$ and
a private rate $R_p=\sum_{k=1}^K R_k$, where $R_k$ denotes the
private rate of the $k$th user. The common rate represents the
contribution of the common stream, whereas the private rate takes
into account all private streams. In general for the instantaneous
rate, we have $ S_r=R_c+R_p$.

In contrast to conventional RS in MISO systems, the $k$th receiver
in a MIMO system has a total of $M_k$ copies of the common symbol
available. These copies can be used to enhance the common rate of
the system. In this section, we propose combining strategies to
improve the common rate performance and also derive an expression
for the private rate.

\subsection{Min-Max Criterion}

Let us consider \eqref{Received signal RS antenna k} from the model
described in section II. The $k$th user receives $M_k$ streams each
one containing a copy of the common symbol. When decoding the common
stream, all private messages are considered additional noise. The
common rate of the $i$th stream intended for user $k$ can be
computed by
\begin{equation}
R_{c,k,i}=\log_2\left(1+\frac{a_c^2\lvert\mathbf{f}_i^H\mathbf{p}_c\rvert^2}{\sum_{j=1}^B a_j^2\lvert\mathbf{f}_i^H\mathbf{p}_j\rvert^2+\lVert\mathbf{g}_i\rVert^2\sigma_n^2}\right).\label{individual Rc minmax criterion}
\end{equation}
It is important to note that in equation \eqref{individual Rc minmax
criterion} we consider an imperfect CSIT scenario, i.e.,
$\mathbf{f_i}=\mathbf{\widehat{f}}_i+\mathbf{\widetilde{f}}_i$. The
error in the channel estimate modelled by $\mathbf{\widetilde{f}}_i$
originates MUI, which limits the overall performance of the system.
Moreover, \eqref{individual Rc minmax criterion}  be used in a
perfect CSIT scenario with $\mathbf{f}^H$ reduced to
$\mathbf{\widehat{f}}^H_i$.

When employing the Min-Max criterion, the $k$th receiver picks the
stream in $\mathcal{B}_k$ that leads to the highest achievable
common rate. This is possible because we assume perfect CSI
available at the receiver. Mathematically, this can be expressed by
\begin{equation}
R_{c,k}^{\left(\text{max}\right)}=\max_{i \in \mathcal{B}_k}\left(R_{c,k,i}\right).\label{MinMax criterion Max}
\end{equation}

Let us consider the vector $\mathbf{r}_c^{\left(\text{max}\right)}$,
containing all the maximums rates from all users i.e. containing all
the rates $R_{c,k}^{\left(\text{max}\right)}$ computed. The common
stream should be decoded by all users. In order to satisfy this
condition we set the common rate equal to the minimum rate stored in
$\mathbf{r}_c^{\left(\text{max}\right)}$, which leads us to
\begin{equation}
R_c^{\left(\text{min-max}\right)}=\min \mathbf{r}_c^{\left(\text{max}\right)}.\label{MinMax criterion Min}
\end{equation}
Finally, the receiver decodes and subtracts  the common stream from
the received signal using SIC in order to decode the private stream.

\subsection{Maximum Rate Combining}

Another possibility to enhance the common rate by exploiting the
multiple streams at the receiver is to use the maximum rate
combining (MRC). Let us consider the received vector of
\eqref{Received signal RS user k}
 and define the combined signal $\tilde{y}_k=\mathbf{w}^H\mathbf{y}_k$, where the vector $\mathbf{w}=\left[w_1~~w_2~~\cdots~~w_{M_k}\right]^{\text{T}}$ represents the combining filter used to maximize the SNR. Then, the average power of $\tilde{y}_k$ is


\begin{align}
\mathbb{E}\left[\lvert\tilde{y}_k\rvert^2\right]=&a_c^2\lvert\boldsymbol{\omega}^{\text{H}}_k\mathbf{H}_k\mathbf{p}_c\rvert^2+\sum_{i\in\mathcal{B}_k}a_i^2\lvert\boldsymbol{\omega}^{\text{H}}_k\mathbf{H}_k\mathbf{p}_i\rvert^2\nonumber\\
&+\sum\limits_{\substack{j=1\\j\notin \mathcal{B}_k}}^B a_j^2\lvert\boldsymbol{\omega}^{\text{H}}_k\mathbf{H}_k\mathbf{p}_j\rvert^2+\mathbb{E}\left[\lvert\boldsymbol{\omega}^{\text{H}}_k\mathbf{n}_k\rvert^2\right],\label{MRC receive power}
\end{align}
where we introduce the row vector $\boldsymbol{\omega}^{\text{H}}_k=\mathbf{w}^{H}\mathbf{G}_k$ in order to simplify the notation. By evaluating the noise term we obtain
\begin{equation}
\mathbb{E}\left[\lvert\boldsymbol{\omega}^{\text{H}}_k\mathbf{n}_k\rvert^2\right]=\lVert\boldsymbol{\omega}_k\rVert^2\sigma_n^2.
\end{equation}

Let us also define the common and private vectors $\mathbf{r}_{k,c}=\mathbf{H}_k\mathbf{p}_c$ and $\mathbf{r}_{k,i}=\mathbf{H}_k\mathbf{p}_i$ with $i \in\mathcal{B}_k$. Substituting these terms in \eqref{MRC receive power} we get
\begin{align}
\mathbb{E}\left[\lvert\tilde{y}_k\rvert^2\right]=&a_c^2\lvert\boldsymbol{\omega}_k^{H} \mathbf{r}_{k,c}\rvert^2+\sum_{i\in\mathcal{B}_k}a_i^2\lvert\boldsymbol{\omega}_k^{H}\mathbf{r}_{k,i}\rvert^2\nonumber\\
&+\sum\limits_{\substack{j=1\\j\notin \mathcal{B}_k}}^B a_j^2\lvert\boldsymbol{\omega}^{H}_k\mathbf{r}_{k,j}\rvert^2+\lVert\boldsymbol{\omega}_k\rVert^2\sigma_n^2.
\end{align}

From the last equation we obtain the SINR for the common message, which is given by

\begin{equation}
\gamma_{k,c}^{\left(\text{MRC}\right)}=\frac{a_c^2\lvert\boldsymbol{\omega}_k^{H}\mathbf{r}_{k,c}\rvert^2}{\sum\limits_{i\in\mathcal{B}_k}a_i^2\lvert\boldsymbol{\omega}_k^{H} \mathbf{r}_{k,i}\rvert^2+\sum\limits_{\substack{j=1\\j\notin \mathcal{B}_k}}^B a_j^2\lvert\boldsymbol{\omega}_k^H\mathbf{r}_{k,j}\rvert^2+\lVert\boldsymbol{\omega}_k\rVert^2\sigma_n^2}.
\end{equation}

Using the property of the dot product and simplifying terms, the SINR can be expressed as follows:
\begin{equation}
\gamma_{k,c}^{\left(\text{MRC}\right)}=\frac{a_c^2\lVert \mathbf{r}_{k,c}\rVert^2\cos\theta}{\sum\limits_{i\in\mathcal{B}_k}a_i^2\lVert \mathbf{r}_{k,i}\rVert^2\cos\beta_i+\sum\limits_{\substack{j=1\\j\notin \mathcal{B}_k}}^B a_j^2\lVert \mathbf{r}_{k,j}\rVert^2\cos\beta_j+\sigma_n^2}
\end{equation}
where $\theta$ is the angle between the vectors $\boldsymbol{\omega}_k$ and $\mathbf{r}_{k,c}$ and $\beta_j$ is the angle between $\boldsymbol{\omega}_k$ and $\mathbf{r}_{k,j}$.

The maximum value of the numerator is achieved when $\cos\theta=1$ and is obtained when the vectors $\boldsymbol{\omega}_k$ and $\mathbf{r}_{k,c}$ are parallel. By setting $\mathbf{w}=\left(\mathbf{G}_k^H\right)^{-1}\frac{\mathbf{r}_{k,c}}{\lVert \mathbf{r}_{k,c}\rVert^2} $ the vectors $\boldsymbol{\omega}_k$ and $\mathbf{r}_{k,c}$ become parallel which lead us to the following SINR expression:

\begin{equation}
\gamma_{k,c}^{\left(\text{MRC}\right)}=\frac{a_c^2\lVert \mathbf{r}_{k,c}\rVert^2}{\sum\limits_{i\in\mathcal{B}_k}a_i^2\lVert \mathbf{r}_{k,i}\rVert^2\cos\beta_i+\sum\limits_{\substack{j=1\\j\notin \mathcal{B}_k}}^B a_j^2\lVert \mathbf{r}_{k,j}\rVert^2\cos\beta_j+\sigma_n^2}.\label{SINR_MRC}
\end{equation}
Since all users should decode the common message, the transmitter set the common rate equal to the minimum rate found across all users i.e., the common rate is given by
\begin{equation}
R_c^{\left(\text{MRC}\right)}=\min_{k=1\cdots K} \log_2\left(1+\gamma_{k,c}^{\left(\text{MRC}\right)}\right)
\end{equation}

\subsection{Private Rate}
After decoding the common stream, the system performs SIC to remove the common symbol from the received signal. Considering that the precoder reduces the interference to the noise level we
have that the covariance matrix of the effective noise is given by
\begin{equation}
\mathbf{R}_{\mathbf{z}_k\mathbf{z}_k}=\sum\limits_{\substack{i=1\\i\neq k}}^K\mathbf{G}_k\mathbf{H}_k\mathbf{P}_i\text{diag}\left(\mathbf{a}_i\odot\mathbf{a}_i\right)\mathbf{P}_i^{H}\mathbf{H}_k^{H}\mathbf{G}_k^{H}+\mathbf{R_{nn}}
\end{equation}
Then, the achievable rate for the  $k$th user is \cite{Cover2006}
\begin{equation}
R_k=\log_2\left(\det\left[\mathbf{I}+\mathbf{F}^H_k\mathbf{P}_k\text{diag}\left(\mathbf{a}_k\odot\mathbf{a}_k\right)\mathbf{P}_k^{H}\mathbf{F}_k\mathbf{R}_{\mathbf{z}_k\mathbf{z}_k}^{-1}\right]\right)\label{Mimo rate}
\end{equation}

\section{Rate Analysis}
In this section, we carry out the sum rate analysis of the proposed strategies combined with the BD precoder.

\subsection{RS Min-Max Criterion with the BD precoder}
The BD precoder partially removes the MUI interference. However, residual interference remains due to the imperfect CSIT, which lead us to the following received vector:
\begin{align}
\mathbf{y}_k=&a_c s_c \underaccent{\ddot}{\mathbf{U}}_k^H\mathbf{H}_k\mathbf{p}_c+\underaccent{\ddot}{\mathbf{\Psi}}_k\text{diag}\left(\mathbf{a}_k\right)\mathbf{s}_k+\mathbf{\tilde{T}}^{\left(k,k\right)}\text{diag}\left(\mathbf{a}_k\right)\mathbf{s}_k\nonumber\\
&+\sum\limits_{\substack{j=1\\j \neq k}}^K \mathbf{\tilde{T}}^{\left(k,j\right)}\text{diag}\left(\mathbf{a}_j\right)\mathbf{s}_j+\underaccent{\ddot}{\mathbf{U}}_k^H\mathbf{n}_k,
\end{align}
where the matrix $\mathbf{\tilde{T}}^{\left(k,j\right)}=\underaccent{\ddot}{\mathbf{U}}_k^H\mathbf{\tilde{H}}_k\mathbf{\bar{V}}_j^{\left(0\right)}\underaccent{\ddot}{\mathbf{V}}_j^{\left(1\right)}$ represents the residual interference. Let us also consider the index $n_k=\sum_{l=1}^{k-1}M_l$. Evaluating the expected value of the received signal power we can obtain
the common rate of the $i$th stream intended for user $k$, which is given
by
\begin{equation}
R_{c,k,i}^{\left(BD\right)}=\log_2\left(1+\frac{a_c^2\lvert\mathbf{f}_i^H\mathbf{p}_c \rvert^2}{\rho^{\left(\text{M-M}\right)}_{k,i}+\sigma_n^2}\right),
\end{equation}
where
\begin{align}
\rho_{k,i}^{\left(\text{M-M}\right)}=&a_i^2\lvert\underaccent{\ddot}{\psi}_{i-n_k}^{\left(k\right)}+\tilde{t}^{\left(k,k\right)}_{i-n_k,i-n_k}\rvert^2+\sum\limits_{\substack{l=1\\l\neq i-n_k}}^{M_k}a_{l+n_k}^2\lvert\tilde{t}^{\left(k,k\right)}_{i-n_k,l}\rvert^2\nonumber\\
&+\sum\limits_{\substack{j=1\\j\neq k}}^K\sum\limits_{q=1}^{M_j}a_{n_j+q}^2\lvert\tilde{t}^{\left(k,j\right)}_{i-n_k,q}\rvert^2
\end{align}

In a perfect CSIT scenario, the precoder and the receiver remove completely the interference and the previous equation is reduced to
\begin{equation}
R_{c,k,i}^{\left(BD\right)}=\log_2\left(1+\frac{a_c^2\lvert\mathbf{f}_i^H\mathbf{p}_c \rvert^2}{a_i^2\lvert\underaccent{\ddot}{\psi}_{i-n_k}^{\left(k\right)}\rvert^2+\sigma_n^2}\right)
\end{equation}

\subsection{RS MRC criterion with the BD precoder}
Let us consider the $k$th user and evaluate the vector $\mathbf{r}_{k,j}$ with $j \in \mathcal{B}_q$. We also define the column index $m=j-\sum_{t=1}^{q-1}M_t$. When $q=k$ the squared module of vector $\mathbf{r}_{k,j}$ is reduced to:
\begin{equation}
\lVert\mathbf{r}_{k,j}\rVert^2=\lVert\underaccent{\ddot}\psi_m^{\left(k\right)}\underaccent{\ddot}{\mathbf{u}}^{\left(k\right)}_m+\underaccent{\ddot}{\mathbf{\tilde{H}}}_k\underaccent{\ddot}{\mathbf{v}}^{\left(k\right)\left(1\right)}_{m}\rVert^2.\label{rdata}
\end{equation}
The BD precoder should reduce the vector $\mathbf{r}_{k,j}$ to zero when $q\neq  k$ due to the zero inter-user interference restriction imposed. However, the imperfect CSIT assumption originates residual MUI, which leads us to
\begin{equation}
\lVert\mathbf{r}_{k,j}\rVert^2=\sum_{i=1}^{N_k}\left\lvert\sum_{l=1}^{N_t}\sum_{n=1}^{N_t}\tilde{h}_{i,l}^{\left(k\right)}\bar{v}_{l,n}^{\left(q\right)\left(0\right)}\underaccent{\ddot}{v}_{n,m}^{\left(q\right)\left(1\right)}\right\rvert^2,\label{rinterference}
\end{equation}


Substituting \eqref{rdata} and \eqref{rinterference} in \eqref{SINR_MRC}  we get the SINR expression, which is given by
\begin{equation}
\gamma_{k,c}^{\left(\text{MRC}\right)}=\frac{a_c^2\lVert \mathbf{r}_{k,c}\rVert^2}{\rho_k^{\left(\text{MRC}\right)}+\sigma_n^2},\label{SINR_MRC_BD}
\end{equation}
where
\begin{align}
\rho_{k,i}^{\left(\text{MRC}\right)}=&\sum\limits_{i\in\mathcal{B}_k}a_i^2\lVert\underaccent{\ddot}\psi_m^{\left(k\right)}\underaccent{\ddot}{\mathbf{u}}^{\left(k\right)}_m+\underaccent{\ddot}{\mathbf{\tilde{H}}}_k\underaccent{\ddot}{\mathbf{v}}^{\left(k\right)\left(1\right)}_{m}\rVert^2\cos\beta_i\nonumber\\&+\sum\limits_{\substack{j=1\\j\notin \mathcal{B}_k}}^B a_j^2\left[\sum_{i=1}^{N_k}\left\lvert\sum_{l=1}^{N_t}\sum_{n=1}^{N_t}\tilde{h}_{i,l}^{\left(k\right)}\bar{v}_{l,n}^{\left(q\right)\left(0\right)}\underaccent{\ddot}{v}_{n,m}^{\left(q\right)\left(1\right)}\right\rvert^2\right]\cos\beta_j.
\end{align}
Under perfect CSIT assumption the MUI interference, which is given by $\underaccent{\ddot}{\mathbf{\tilde{H}}}_k\underaccent{\ddot}{\mathbf{v}}^{\left(k\right)\left(1\right)}_{m}$ and $\lVert \mathbf{r}_{k,j} \rVert^2$ with $j\notin \mathcal{B}_k$, is eliminated and the expression in \eqref{SINR_MRC_BD} is reduced to
\begin{equation}
\gamma_{k,c}^{\left(\text{MRC}\right)}=\frac{a_c^2\lVert \mathbf{r}_{k,c}\rVert^2}{\sum\limits_{i \in \mathcal{B}_k}a_i^2\lvert\underaccent{\ddot}{\psi}_{i-n_k}^{\left(k\right)}\rvert^2\cos\beta_i+\sigma_n^2}.
\end{equation}

\subsection{RS BD private streams sum rate}
Let us consider the matrix $\mathbf{\Phi_k}=\underaccent{\ddot}{\mathbf{\Psi}}_k+\mathbf{\tilde{F}}_k$. After the SIC process we get the equations for the private rate
\begin{equation}
R_p^{\left(\text{BD}\right)}=\log_2\det\left[\mathbf{I}+\mathbf{\Phi_k}\text{diag}\left(\mathbf{a}_k\odot\mathbf{a}_k\right)\mathbf{\Phi_k}^{H}\mathbf{R}_{\mathbf{z}_k\mathbf{z}_k}^{-1}\right].
\end{equation}
Under perfect CSIT assumption we have
\begin{equation}
R_p^{\left(\text{BD}\right)}=\log_2\det\left[\mathbf{I}+\underaccent{\ddot}{\mathbf{\Psi}}_k\text{diag}\left(\mathbf{a}_k\odot\mathbf{a}_k\right)\underaccent{\ddot}{\mathbf{\Psi}}_k^{H}\mathbf{R}_{\mathbf{n}\mathbf{n}}^{-1}\right].
\end{equation}

\section{Simulations}

In this section we assess the performance of the proposed MIMO RS
schemes employing ZF and BD precoders. A total of 12 transmit
antennas was used at the BS for all simulations. We also consider 6
users where each is equipped with 2 receive antennas. The inputs follow a
Gaussian distribution with zero mean and variance equal to one. We
consider additive white Gaussian noise with the same statistics for
all users, such that all users experience the same SNR. The ESR was
computed averaging 100 independent channel realizations. For each
channel realization we obtained the ASR employing 100 error
matrices. The common precoder was set to
$\mathbf{p}_c=\mathbf{V}\left(:,1\right)$ where we use a singular
value decomposition of the channel matrix
$\left(\mathbf{H}=\mathbf{U\Psi V}\right)$. The power allocated to the
common precoder was found through exhaustive search, where we keep
the proportion of power allocated to the private precoders fixed.
Conventional RS, which uses the minimum common rate available, was
also considered. We termed this strategy as RS in the simulation
results.

\begin{figure}[htb]
\begin{centering}
\includegraphics[scale=0.4]{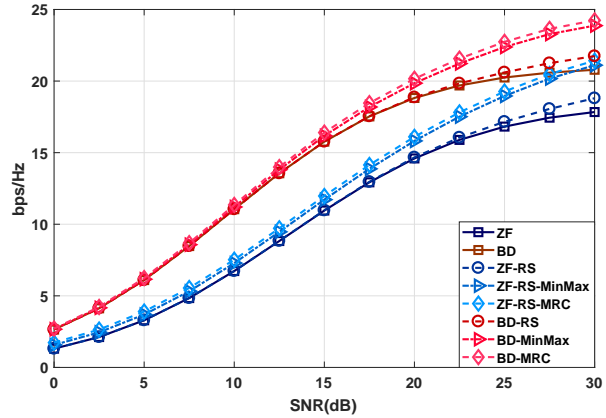}
\par\end{centering} \vspace{-0.75em}
\caption{Sum-rate performance with imperfect CSIT and fixed error variance.}\label{SumRateZF}
\end{figure}

Fig. \ref{SumRateZF} summarizes the sum rate performance of the ZF precoder and the BD precoder, both operating in a MU-MIMO system. For this simulation we consider a fixed error variance in the channel equal to 0.1.  The proposed techniques achieve a better results because they exploit the multiple antennas at the receiver, enhancing the common stream. The BD precoder allows not only the enhancement of the common stream but also of the private stream obtaining a better performance in terms of sum rate. The best performance is achieved by the BD-RS-MRC due to the use and combination of all available signals at the receive antennas. The curves obtained exhibit saturation because of the imperfect CSIT assumption, which originates MUI that scales with the SNR.

\begin{figure}[htb]
\begin{centering}
\includegraphics[scale=0.4]{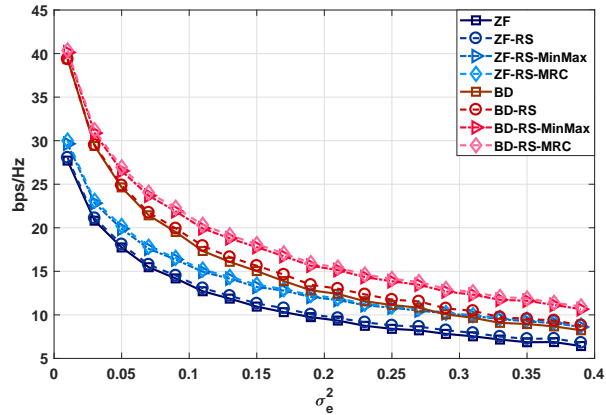}
\par\end{centering} \vspace{-0.75em}
\caption{Error variance VS Sum Rate performance.}\label{VarErrZF}
\end{figure}

For the second scenario we evaluate the performance of the proposed
schemes operating at different noise levels as depicted in Fig. \ref{VarErrZF}
The SNR was set to 25 dB. The results show that
the RS strategies increase the robustness of the system across all
error variances. The proposed MIMO BD-RS-MRC strategy achieves the
highest sum-rate, which is up to 35\% higher when compared to the
conventional BD precoder.

\begin{figure}[htb]
\begin{centering}
\includegraphics[scale=0.4]{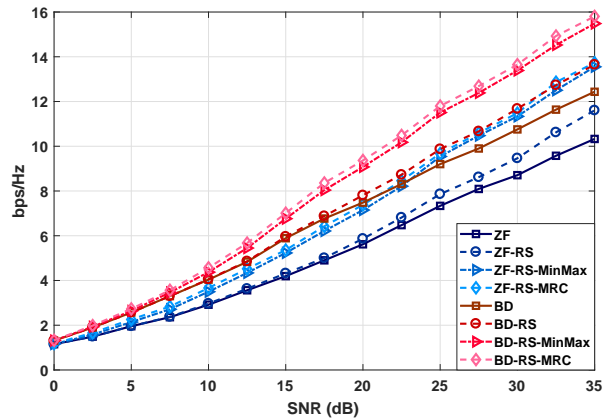}
\par\end{centering} \vspace{-0.75em}
\caption{Sum-rate performance with imperfect CSIT and varying channel estimate errors.}\label{QualityZF}
\end{figure}

For the last example, we consider that the quality of the channel estimate
improves with the SNR, i.e. $\sigma^2_e=\beta
\left(E_{tr}/\sigma^2_n\right)^{-\alpha}$. The parameters $\beta$
and $\alpha$ were set to 0.94 and 0.2 respectively. Fig.
\ref{QualityZF} shows that the proposed schemes are more
robust than conventional precoding schemes. MIMO BD-RS-MRC shows a sum
rate improvement of 33,33\% when compared to conventional BD precoding,
whereas the MIMO ZF-RS-MRC achieves a sum rate 35\% higher than the
conventional ZF-strategy.

\section{Conclusion}
In this paper, we have proposed MIMO RS strategies combined with the BD precoder and two criteria to enhance the common rate by taking advantage of the multiple antennas at the receivers. In general, all BD precoder schemes outperform their ZF precoder counterpart in terms of sum rate. Moreover, the BD-RS-MRC scheme achieves the best performance among all the proposed techniques, attaining an improvement of more than 30\% when compared to conventional techniques. Simulation results have also shown that the BD-RS scheme is more robust when compared to ZF techniques under imperfect CSIT scenarios.
\bibliographystyle{IEEEtran}
\bibliography{PrecodingAlgorihtms}
%

%
%

\end{document}